# Experimental and Computational Analysis of Energy Absorption Characteristics of Three Biomimetic Lattice Structures Under Compression

Mahtab Vafaeefar[1], Kevin M. Moerman[2,3, *], Ted J. Vaughan[1, *]

[1] Biomechanics Research Centre (BMEC), School of Engineering, College of Science and Engineering, University of Galway, Ireland.
[2] Mechanical Engineering, School of Engineering, College of Science and Engineering, University of Galway, Ireland.
[3] Griffith Centre of Biomedical and Rehabilitation Engineering (GCORE), Griffith University, Gold Coast, Australia.

*Correspondence: Authors contributed equally and are co-senior/corresponding authors*

**Dr Ted Vaughan:**     ted.vaughan@universityofgalway.ie

**Dr Kevin Moerman:**     kevin.moerman@universityofgalway.ie

**Keywords**
Energy absorption, Dual-lattice, Gyroid, Lattices, Computational modelling, Densification

Abstract

The objective of this study is to evaluate the mechanical properties and energy absorption characteristics of the gyroid, dual-lattice and spinodoid structures, as biomimetic lattices, through finite element analysis and experimental characterisation. As part of the study, gyroid and dual-lattice structures at 10% volume fraction were 3D-printed using an elastic resin, and mechanically tested under uniaxial compression. Computational models were calibrated to the observed experimental data and the response of higher volume fraction structures were simulated in an explicit finite element solver. Stress-strain data of groups of lattices at different volume fractions were studied and energy absorption parameters including total energy absorbed per unit volume, energy absorption efficiency and onset of densification strain were calculated. Also, the structures were characterized into bending-dominant and stretch-dominant structures, according to their nodal connectivity and Gibson-and-Ashby's law. The results of the study showed that the dual-lattice is capable of absorbing more energy at each volume fraction cohort. However, gyroid structures showed higher energy absorption efficiency and the onset of densification at higher strains. The spinodoid structure was found to be the poorest structure in terms of energy absorption, specifically at low volume fractions. Also, the results showed that the dual-lattice was a stretch dominated structure, while the gyroid structure was a bending dominated structure, which may be a reason that it is a better candidate for energy absorption applications.

# 1 Introduction

Porous engineering structures, such as lattices and foams, offer a combination of tailored mechanical properties such as high strength, energy absorption and light weight, and have become increasingly popular for multifunctional applications (Bian et al., 2021; Ha et al., 2019; Maskery et al., 2017; Yin et al., 2023). Lattice structures are of particular importance for energy absorption applications, such as in packaging materials, personal protective equipment, automotive materials, and military equipment (Maskery et al., 2017; Mueller et al., 2019). With the advent of additive manufacturing techniques, the creation of sophisticated lattice structures has been revolutionized, making it possible to fabricate highly tortuous and controlled structures (Afshar et al., 2018).

Energy absorption in lattice structures can be defined as their ability to transform kinetic energy into other forms of energy through elastic/plastic deformation, mechanical instabilities, and failure (Habib et al., 2017; Rahman et al., 2021). Enhanced energy absorption in porous structures is achieved by reducing the peak stress throughout the structure under high deformations, and by promoting an even distribution of stress within the material upon loading (Mueller et al., 2019). These porous structures are capable of large deformations at almost constant stress levels, which makes them an ideal candidate for energy absorption applications (Habib et al., 2018). Many of the current applications of these energy absorbent lattices involve stochastic foams or conventional honeycomb and sandwich panel structures, which are not optimised for this function (Mueller et al., 2019; Rahman et al., 2021). Therefore, to address the issue, new lattice designs with enhanced mechanical characteristics and energy absorption capacity have been inspired by natural cellular structures, derived from plants and animals. It has been shown that these biomimetic lattice structures are viable alternatives that offer high energy absorption for multifunctional applications (Rahman et al., 2021). Examples of these biomimetic approaches include thin-wall bamboo structures (Fu et al., 2019; Ufodike et al., 2021), bone (Xiang et al., 2020; Zhang et al., 2022) and tendon (Tarafdar et al., 2021; Tsang & Raza, 2018), biomimetic sandwich panels and lattices (Ha & Lu, 2020) with excellent energy absorption capacity.

The trabecular bone structure, which is comprised of interconnected plate and rod elements, is a tailored microarchitecture for load-bearing capability. Hence, various biomimetic lattices have been proposed to imitate the features of this bone structure. Early computational structures have been proposed to depict trabecular bone microarchitecture in the form of biomimetic lattices (Liu et al., 2006; Rammohan et al., 2015; Yin et al., 2023; X. Y. Zhang et al., 2020). Some of these studies have used repeating-unit lattices (Colabella et al., 2017; Gibson, 2005, 2018). While these provided significant knowledge on the mechanics of bone structure, the non-uniformity and general complexity in the trabecular bone microarchitecture are not adequately represented by these repeating-unit lattices.

In a recent study (Vafaeefar et al., 2023), we investigated three structures as biomimetic lattices of trabecular bone. These three structures include the gyroid, as a periodic lattice, the stochastic spinodoid structure and the newly introduced dual-lattice structure. The gyroid, a member of the triply periodic minimal surface (TPMS) family, enables the creation of smooth, uninterrupted surfaces with adjustable mechanical characteristics (Peng & Tran, 2020). The surface-based gyroid structure has been studied for multiple applications, including energy absorbance (Lu et al., 2021; Maskery et al., 2017; Wang et al., 2020). The gyroid structure offers parametric design capabilities and has been fabricated with different additive manufacturing techniques (Ataee et al., 2018; Lu et al., 2021; Maskery et al., 2017). The gyroid structure has been analysed as energy absorbent cellular structures (D. Li et al., 2019; Yang et al., 2019; Yu et al., 2019; X. Y. Zhang et al., 2020). However, in previous studies, due to the brittleness of the parent material in manufacturing, deformation was accompanied with fracture and failure, and poor energy dissipation (Andrew et al., 2021; Z. Fan et al., 2022; Maskery et al., 2017). Irregular lattice structures, like the Voronoi-based lattices (Kirby et al., 2020; Ruiz et al., 2010, 2011; Silva & Gibson, 1997), have been studied recently and tried to capture the complex geometry of trabecular bone. Gaussian random field functions (Padilla et al., 2008) and spinodoid structures (Kumar et al., 2020), are stochastic structures, that have recently been studied as trabecular bone mimicking lattices (Callens et al., 2021). The microarchitectural characteristics of the spinodoid structures (Kwon et al., 2009; Vafaeefar et al., 2023) and its elastic mechanical properties (Kumar et al., 2020; Soyarslan et al., 2018) have been studied in detail previously, but there is limited understanding of its energy absorption capability. Recently, we developed a new dual-lattice structure, which is a Voronoi-like bone mimetic lattice structure, and its elastic mechanical capabilities besides its microarchitectural features compared to trabecular bone were studied (Vafaeefar et al., 2023). While we demonstrated that these structures have similar performance to trabecular bone structures in the elastic regime, the energy absorption characteristics (Kansara et al., 2021) of these structures have never been explored.

In the present study, the mechanical properties and energy absorption characteristics of the well-known gyroid structure, the newly introduced dual-lattice structure (Vafaeefar et al., 2023) and the stochastic spinodoid structure, under high deformations were investigated numerically and experimentally. The three structures were generated as biomimetic lattices inspired by trabecular bone microarchitecture. A hyperelastic material was used to address the research question in this study, that how these structures would behave under high deformations. Finite element (FE) models of a range of volume fractions (10%, 20%, 30%, 40%, and 50%) of the three structure types were generated and analysed. Material calibration was performed using experimental data for uniaxial compression test on 3D printed sample of the two 10% volume fractions of the gyroid and dual-lattice structures. The mechanical properties, energy absorption, and energy efficiency characteristics of three lattice types were studied. Different volume fractions of the structures were considered, and their stress-strain response were analysed. With variable volume fractions, the effect of structural indices on their energy absorption response were investigated.

## 2 Materials and methods

### 2.1 Structure design

The three lattices considered in this study are Figure 1the gyroid structure (Figure 1 (a)) (Rammohan et al., 2015; Yang et al., 2019), the dual-lattice (Vafaeefar et al., 2023) (Figure 1 (b)), and the stochastic spinodoid structure (Figure 1 (c)) (Kumar et al., 2020). For each structure, the input variables were selected such that the three structures have the same volume fractions. All the structures were implemented using the open-source MATLAB toolbox GIBBON (Moerman, 2018).

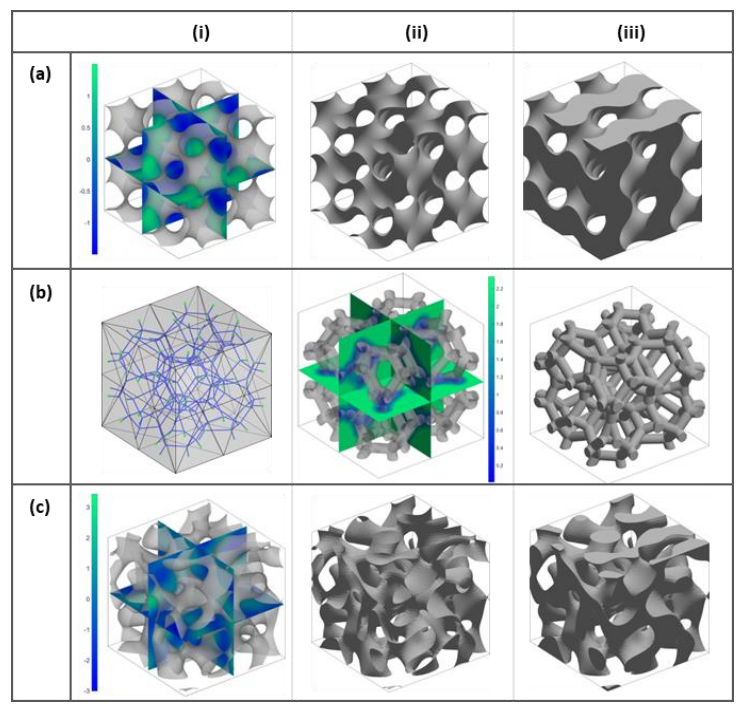

Figure 1: Models of the gyroid (a), dual-lattice (b), and spinodoid (c) structures, and illustration of the steps to generate each structure (I -III).

### 2.1.1 Gyroid structure

TPMS structures are the most common type of surface-based lattice structures and exhibit a zero mean curvature (Echeta et al., 2020). Out of the different structures of the TPMS family, this study focusses on the gyroid, which is described by the following level-set function $f$ (C. Zhang et al., 2022):

$$f(x,y,z) = \left[\sin\left(\frac{2\pi}{a_x}x\right)\cos\left(\frac{2\pi}{a_y}y\right)\right] + \left[\sin\left(\frac{2\pi}{a_y}y\right)\cos\left(\frac{2\pi}{a_z}z\right)\right]$$
$$+ \left[\sin\left(\frac{2\pi}{a_z}z\right)\cos\left(\frac{2\pi}{a_x}x\right)\right], \quad (1)$$

where $a_i$, ($i = x, y, z$) is the wavelength, representing number of unit cells in each direction in the lattice structure. By introducing a level-set value $f_0$, an offset surface is reconstructed from the minimal surface $f = 0$. This level-set value binarizes the domain into two bi-continuous volumes (Mangipudi et al., 2016) by generating the level-set surfaces (Figure 1 (a-i)), and generates the shell-type gyroid (Figure 1 (a-ii)) (Callens et al., 2021). The solid-type gyroid by closing-over the boundaries and solidifying the closed surface for $f \geq f_0$ was created for this study. Varying the parameter $f_0$, results in varying the surface offsets, and results in different volume fractions of the gyroid structure. By closing the caps of the solid section in between the sheets, the final solid gyroid structure (Figure 1 (a-iii)) was created.

*2.1.2 Dual-Lattice structure*

The dual-lattice structure is based on thickening the struts of the dual of a Delaunay tessellation, and the process of generating a dual-lattice structure is shown in Figure 1. In the first step (Figure 1 (b-i)), the domain is discretised into tetrahedral elements using the TetGen algorithm (Si, 2015) by introducing the point spacing and stretch factor parameters, which control the size of the cells and anisotropy in the structure, respectively. A skeleton model is generated by connecting the centre of each cell to those of its adjacent cells (or face centre for boundary tetrahedra) (Figure 1 (b-i)). Next, a type of level-set image is defined by computing the distance of the skeleton to a surrounding image grid. A thickened version of the skeleton is then obtained by computing the level-set surface at a desired distance from the skeleton (Figure 1 (b-ii)). The obtained surface is then closed which results in the final dual-lattice structure (Figure 1 (b-iii)).

*2.1.3 Spinodoid structure*

The spinodoid is a stochastic foam, generated based on a phase-separation process using a Gaussian Random Field (GRF) function (Cahn, 1965). The structure is generated by superimposing many standing waves ($N \gg 1$), of fixed wavelengths (using wave number $\beta$), but random directions ($q_i$) and phases ($\varphi_i$), and can be mathematically described by the following level-set function (Kumar et al., 2020; Soyarslan et al., 2018):

$$f(\mathbf{x}) = \sqrt{\frac{2}{N}}\left[\sum_{i=1}^{N}\cos(\beta q_i \mathbf{x} + \varphi_i)\right] \quad (2)$$

where $\mathbf{x}$ is the position vector. An anisotropic extension, using an orientation distribution function (ODF), can be applied to the GRF function, to generate anisotropic lattice structures by favouring certain directions of waves and neglecting the others in an isotropic environment (Kumar et al., 2020). Similar to the gyroid structure, by introducing the level-set parameter $f_0$, a bi-continuous topology is constructed (Figure 1 (c-i)). The $f_0$ parameter is then defined as the quantile evaluated at the average relative density $\rho$ of the solid phase, and is computed as (Kumar et al., 2020):

$$f_0 = \sqrt{2}\,[\mathrm{erf}^{-1}(2\rho - 1)] \qquad (3)$$

The domain was binarized into solid and void sections (Figure 1 (c-ii)) by the binary indicator function $S(x)$ at position vector **x** over the domain:

$$S(\mathbf{x}) = \begin{cases} 1(solid) & if \quad f(\mathbf{x}) \leq f_0 \\ 0(void) & if \quad f(\mathbf{x}) > f_0 \end{cases} \qquad (4)$$

With the same process of the gyroid structure, by closing the surfaces in the solid area, the final spinodoid structure is generated (Figure 1 (c-iii)).

*2.2 Microarchitecture of lattices*

The mechanical characteristics of lattices, such as their stiffness and strength, rely on the characteristics of the parent material, their relative density, and the architecture of the lattices (Ashby, 2006). For the lattice structures generated, a detailed analysis of the microstructural indices of the structures were performed using BoneJ (Doube et al., 2010). These parameters are specific to trabecular bone; however, they can be used to quantify the microstructural parameters of the lattices in this study. A summary of the parameters considered, and a brief description of each, is presented in Table 1.

Table 1 Summary of studied microstructural and morphometric indices.

| Parameter | Abbreviation, unit | Description |
| --- | --- | --- |
| Volume fraction | VF (%) | The fraction of solid volume to the total cube volume of the structure. |
| Mean trabeculae thickness | Tb.Th (μm) | The average width of the structure's elements. |
| Mean trabeculae spacing | Tb.Sp (μm) | The average distance between adjacent elements within the structure. |

One important topological factor determining the deformation pattern, and dominant deformation mode of lattices, is their strut connectivity, and joints (Kadkhodapour et al., 2014). Based on the topological factors of the structures, they can be bending-dominated structure, meaning that under compression loading, the struts of the frame bend, or a stretch dominated structure, where the members carry tension or compression (Ashby, 2006). A larger ratio of struts to joints, or generally larger nodal connectivity, will result in more stretch-dominance performance (Kadkhodapour et al., 2014). Stretch-dominated lattices are renowned for their high specific stiffness and strength, compared to the same volume fraction of bending-dominated structures (Ashby, 2006; H. L. Fan et al., 2006; Peng et al., 2020). In this study, nodal connectivity was used for the lattice structures to evaluate their deformation dominance behaviour.

Normalized elastic modulus by the solid (parent material) elastic modulus scaling analysis against to the normalized density can be used to define specific mechanical properties of lattices. According to the Ashby's scaling law (Ashby, 2006; Gibson & Ashby, 1997), the normalized Young's modulus varies exponentially as a function of density:

$$(E/E_s) = c(\rho/\rho_s)^n \tag{5}$$

in which $E$ is the Young's modulus of the structure, $E_s$ is the Young's modulus of the solid material that which the structure is made, and $(\rho/\rho_s)$ is the volume fraction of the structure. The constant $c$ is dependent on the parent material and lattice structure. For a wide range of bending-dominated structures, the exponential value $n$ is close to 2, and for the stretch-dominated structures this value approaches 1.

## 2.3 Experimental testing

### 2.3.1 Sample fabrication

To determine the energy absorption under large deformation, a highly elastic material was used to manufacture the lattices. Furthermore, the detail and complexity of the structures demands advanced manufacturing techniques, such as Laser Stereolithography (SLA). Here a Form3 (FormLabs, Somerville, MA, US) SLA 3D printer was used, in combination with the highly flexible "Elastic V50 A" resin (FormLabs, Somerville, MA, US). To avoid undesirable boundary effects, and in anticipation of generating computational models, a 1 mm thick square top and bottom plate were added to the geometry for 3D printing, such that the lattice end features were continuous with the plates. These plates also enable uniform loading of the top and bottom surfaces. Standard Tessellation Language (STL) files of 10% volume fractions of the gyroid and dual-lattice samples (60 mm x 60 mm x 62 mm) were created in MATLAB and imported to PreForm 3.28.0 software (FormLabs, Somerville, MA, US) for pre-processing samples, as shown in Figure 2 (a, b). To address the printability of the structures, supports, including internal supports inside the pores of the structures, were autogenerated in the software with a touchpoint size of 0.55 mm. The samples were printed on a raft and tilted 20° around the $y$ axis to reduce the internal supports. The layer thickness was set to 0.1 mm, resulting in around 800-850 layers per sample.

After printing, samples were washed for 10 minutes in 70% ethanol using a Form Wash (FormLabs, Somerville, MA, US) machine, and all the supports, including the internal supports within the porous structure, and rafts were removed. The samples were post-cured through a multi-directional LED process for 20 minutes at 60°C using a Form Cure (FormLabs, Somerville, MA, US) machine according to the supplier specifications for the elastic V50 A resin. Examples of final 3D printed samples are shown in Figure 2 (c, d).

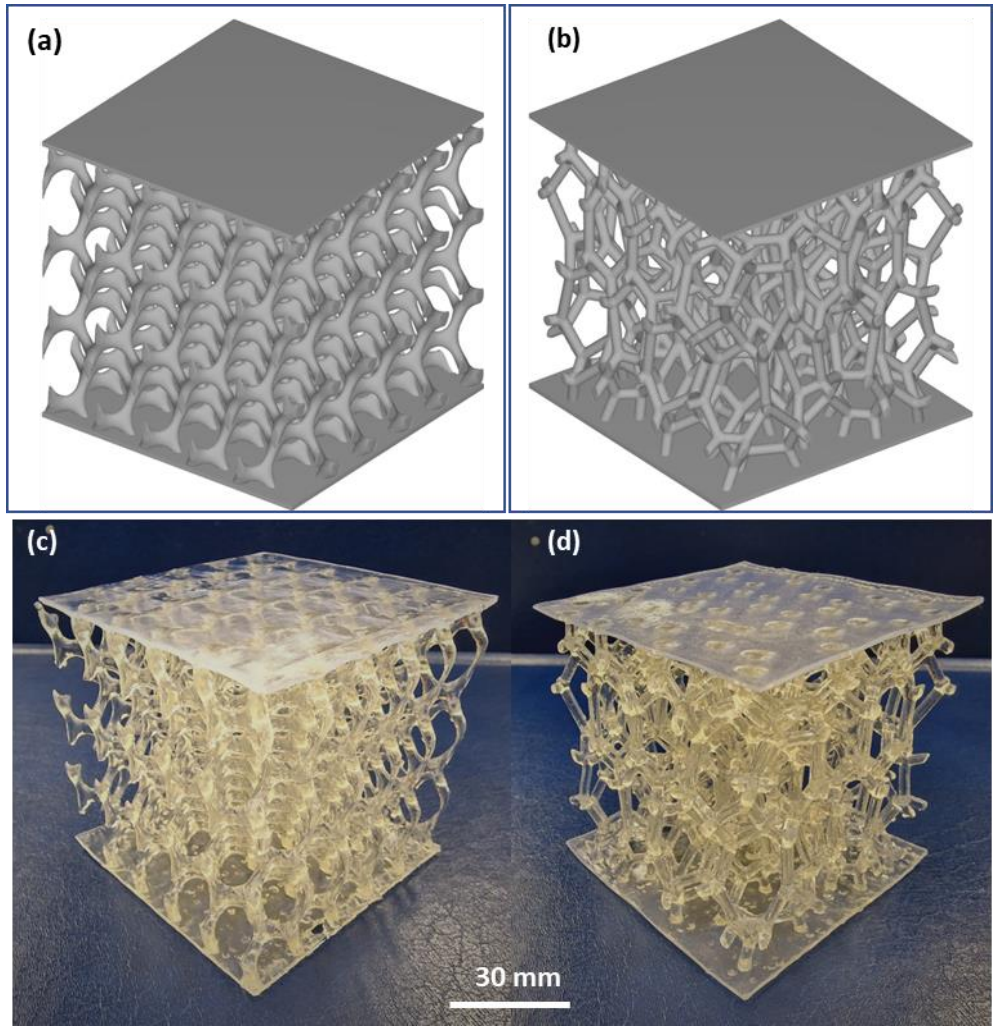

Figure 2: The gyroid (a), and dual-lattice (b) 10% volume fractions structures generated for 3D printing, and the 3D printed samples (c,d)

### 2.3.2 Mechanical testing

To evaluate the mechanical response of the lattice structures, a Zwick mechanical testing machine (Zwick Roell, GmbH & Co., Germany) equipped with a 2.5 kN load cell, was used to perform compressive tests under displacement control at a rate of 5 mm/min (strain rate of $2.6 \times 10^{-3}/s$). Figure 3 shows the mechanical test set-up for compressive tests on the 3D printed samples. Compressive tests were repeated on 3 printed samples for each of the gyroid and dual-lattice structures with 10% volume fraction. Each sample was subjected to a maximum strain of 85% linearly. The data acquisition rate was 10 Hz and force-displacement data was recorded during the test. Engineering stress $\sigma$, and strain $\varepsilon$ were calculated as $\sigma = F/A_{eff}$, where $F$ is the load measured by the load cell and $A_{eff}$ is the original effective cross section area of the lattice structure, and $\varepsilon =$ $\Delta L/L_0$ where $\Delta L$ is the displacement measured by the load cell and $L_0$ is the original height of the sample.

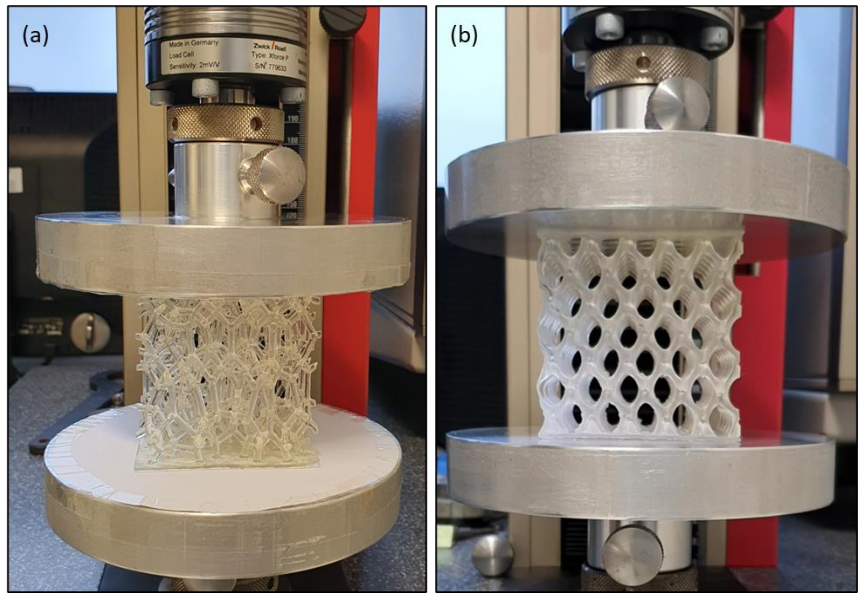

Figure 3: Mechanical test setup for (a) 10% volume fraction of dual-lattice and (b) gyroid 3D printed samples.

## 2.4  Finite Element modelling

### 2.4.1  Geometry and mesh sensitivity study

To study the mechanical properties of the lattice structures, quasi-static finite element analysis was utilised. Figure 4 (a) shows a typical finite element model (FEM) for a gyroid structure. After generating the structures, a high-quality surface triangulation was obtained in MATLAB by using an isotropic surface meshing algorithm (Figure 4 (b)) (Lévy & Bonneel, 2013). The interior was next filled with linear tetrahedral elements (Abaqus element type C3D4) using the TetGen algorithm (Si, 2015). Two flat rigid plates on the top and bottom of the sample were included to enable uniaxial compression to be applied in the FE model. These plates were defined as meshless analytical rigid surfaces.

A mesh sensitivity analysis was carried out to ensure accurate results in the FE analysis. Samples with 10% volume fraction were used for these mesh studies. The gyroid model consisted of 5.33×5.33×3.33 unit cells and the dual-lattice of 4×4×2 unit cells, and both represented 2 mm×2 mm×2 mm cubes. Structures with node spacings of 20 μm, 13 μm and 10 μm were generated and the resulting effective stress-strain curves were determined. The mesh sensitivity analysis showed that mesh size of 13 μm is small enough to guarantees mesh-independent results, and to keep the computational cost manageable with our facilities, in terms of CPU time for analysis and mesh generation. Figure 4 (b) shows the quality and density of the converged mesh size. The mesh sensitivity study is summarized in Table 2, with the results from the study presented in Section 3.1 (see Figure 5).

Table 2 Summary of meshing quality and calculation time

| Model | Mesh Size | Number of Nodes | Number of elements | CPUh |
|---|---|---|---|---|
| GY-10-Ns100 | 20 μm | 186,551 | 771,470 | 03:19:43 |
| GY-10-Ns150 | 13 μm | 536,734 | 2,550,172 | 11:30:14 |
| GY-10-Ns200 | 10 μm | 1,196,150 | 6,104,797 | 22:13:49 |
| DL-10-PS0200 | 20 μm | 168,391 | 681,864 | 03:59:33 |
| DL-10-PS0133 | 13 μm | 537,115 | 2,536,941 | 18:32:54 |
| DL-10-PS0100 | 10 μm | 1,178,995 | 5,980,135 | 28:18:21 |

*2.4.2  Material properties and calibration*

The mechanical properties of the cured resin were already established through uniaxial tensile testing in a former work by Bernini et al. (Bernini et al., 2023). To briefly summarise, the mechanical properties were determined by 3D printing ASTM D638 type IV tensile bars, which were subjected to uniaxial tensile testing using a Test Resources Universal Testing Machine (Actuator, Model SM-500-294 Load Cell, Epsilon Technology Corp. Axial Extensometer Model 3542-0100-050-ST). The current study conducts its own material calibration process of the experimental data (Bernini et al., 2023), whereby hyperelastic material parameters were next determined by fitting the following Neo-Hookean formulation to the experimental stress-strain data. The strain energy function, $U$, is used to characterize the Neo-Hookean material model (ABAQUS Analysis User's Manual, 2014) as follows:

$$U = C_{10}(\overline{I}_1 - 3) + \frac{1}{D_1}(J_e - 1)^2 \quad (6)$$

in which $C_{10}$ and $D_1$ are material constants, relating to shear and volumetric behaviour, respectively. $\overline{I}_1$ is the first invariant of the deviatoric right Cauchy-Green deformation tensor, and $J_e$ is the elastic volume ratio. The $C_{10}$ and $D_1$ were calibrated on the uniaxial compressive response of the gyroid and dual-lattice 10% volume fractions (see Section 2.3.2), while ensuring these parameters also fit the previously measured uniaxial tensile response (Bernini et al., 2023). The final material properties that were used for our FE models, is summarised in Table 3, which are in close agreement with the tensile test data.

Table 3. Mechanical and physical properties of the material used in FEA.

| Material | Neo-Hookean Model | | Density |
|---|---|---|---|
| | $C_{10}$ | $D_1$ | |
| Elastic resin 50A | 0.60 (MPa) | 0.25 (MPa$^{-1}$) | 1.20 g/cm$^3$ |

### 2.4.3 Boundary conditions and model formulation

Figure 4 (a) shows the boundary and loading conditions that were prescribed to the FE model. The top and bottom nodes of the structures were tied to the top and bottom plates, respectively, to avoid detaching from the plates. The bottom plate was fixed in all degrees of freedom. A general contact algorithm was used. A hard normal contact with tangential friction factor of $\mu = 0.6$ was defined for all surfaces in contact including the lattice self-contact. Low results sensitivity was found for the friction factor in the range of 0.6 to 0.9. A vertical displacement up to 1.7 mm ($\varepsilon = 85\%$) in the $z-$direction was applied to the top plate, with all the other degrees of freedom fixed.

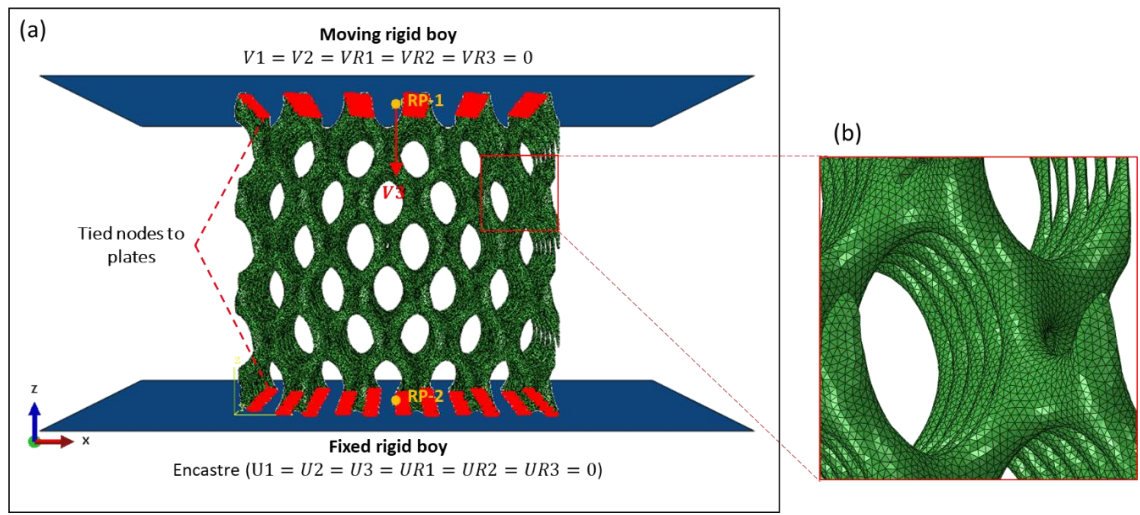

Figure 4: An example visualisation of a 10% volume fraction Gyroid finite element model. The tied boundary conditions are shown in red (a), and a close-up of the triangulated surface mesh is shown in (b).

Due to significant amount of non-linearity in the simulation, an explicit, dynamic solution was used through the Abaqus/Explicit solver (*ABAQUS Finite Element Analysis Software*, 2019) for all the simulations. Moreover, the calculation time was reduced by applying a mass scaling technique. A fixed mass-scaling factor of 1,000 was directly applied to the entire model. To get rate-independent material behaviour, and reduce computation time, the loading rate was adjusted. According to the FE model outputs, applied strain rate fulfilled the requirements for quasi-static analysis, whereby the total kinetic energy of the deforming structure did not exceed 5% of its internal energy throughout most of the process. Simulations were run on a High-Performance Computing cluster, featuring 80 nodes, each node has 2x20-core 2.4 GHz processors and 192 GB of dedicated RAM, and for this study, 2 nodes were used for each simulation.

### 2.4.4 Parameter study

Table 4 shows the input parameters for each structure type to generate different volume fractions. For each structure type, a single independent parameter allows for the variation of the volume fraction. For the gyroid and spinodoid structures, this parameter was the level-set value. By increasing the level-set ($f_0$), the structures thicken, resulting in an increased volume fraction. For the dual-lattice structure, the strut thickness ($d$) was set to a desired value, which in turn directly controls the volume fraction. For all the structures, an anisotropy vector, or unit cell vector was selected to apply anisotropy factor in the direction of load application, here $z$-axis, as reported in Table 4. This anisotropy factor comes from the anisotropic bone structure and making the structure stiffer in one specific direction, according to its application.

Table 4 input parameters for different volume fractions of three types of structures

| Structure Type | Structure design parameters | | | |
|---|---|---|---|---|
| | level-set value, $f_0$ | Unit Cells | Volume Fraction | Sample Size (mm) |
| Gyroid 10% | 1.18 | [5.33, 5.33, 3.33] | 0.1 | 2 |
| Gyroid 20% | 0.91 | [5.33, 5.33, 3.33] | 0.2 | 2 |
| Gyroid 30% | 0.62 | [5.33, 5.33, 3.33] | 0.3 | 2 |
| Gyroid 40% | 0.32 | [5.33, 5.33, 3.33] | 0.4 | 2 |
| Gyroid 50% | 0 | [5.33, 5.33, 3.33] | 0.5 | 2 |
| | level-set value, $f_0$ | Anisotropy Factor | | |
| Spinodiod 30% | 0.3 | [43, 43, 0] | 0.3 | 2 |
| Spinodiod 40% | 0.4 | [43, 43, 0] | 0.4 | 2 |
| Spinodiod 50% | 0.5 | [43, 43, 0] | 0.5 | 2 |
| | strut thickness, $d$ | Unit Cells | | |
| Dual-Lattice 10% | 0.074 | [4, 4, 2] | 0.1 | 2 |
| Dual-Lattice 20% | 0.11 | [4, 4, 2] | 0.2 | 2 |
| Dual-Lattice 30% | 0.14 | [4, 4, 2] | 0.3 | 2 |
| Dual-Lattice 40% | 0.17 | [4, 4, 2] | 0.4 | 2 |
| Dual-Lattice 50% | 0.192 | [4, 4, 2] | 0.5 | 2 |

For all models, the mechanical response of the lattice structure is presented in terms of their compressive stress-strain behaviour. To derive representative nominal stress-strain data, the prescribed top plate displacement, and predicted top plate reaction force were used. The plate force data was converted to nominal stress ($\sigma$) by dividing them to the original cross section area of the FE samples ($A_{eff} = 4$ mm$^2$). The plate displacement was converted to nominal strain ($\varepsilon$) by dividing to the original length of the FE samples ($L_0 = 2$).

## 2.5 Energy absorption characterization of lattices

The stress-strain curves of the lattice structures typically comprise of three main stages, as shown in Figure 5. During the first stage, the structure undergoes an elastic deformation, and often presents with close to linear-elastic behaviour. At a particular stress level, the structure will deform in a nonlinear manner. This nonlinear deformation is typically attributed to elastic buckling, plastic yielding, or fracture (or a combination of these phenomena). During this second stage of deformation, the stress is close-to constant, and is referred to as the plateau stress, $\sigma_{pl}$. The plateau stress is calculated as the average of the stress in the plateau region, which starts at yield stress and ends at the onset of densification (Habib et al., 2018). Once struts have collapsed, the so-called densification stage occurs, in which stress increases rapidly, and the structure starts to behave like the bulk parent material (Gibson, 2005; Habib et al., 2018; Ma et al., 2020).

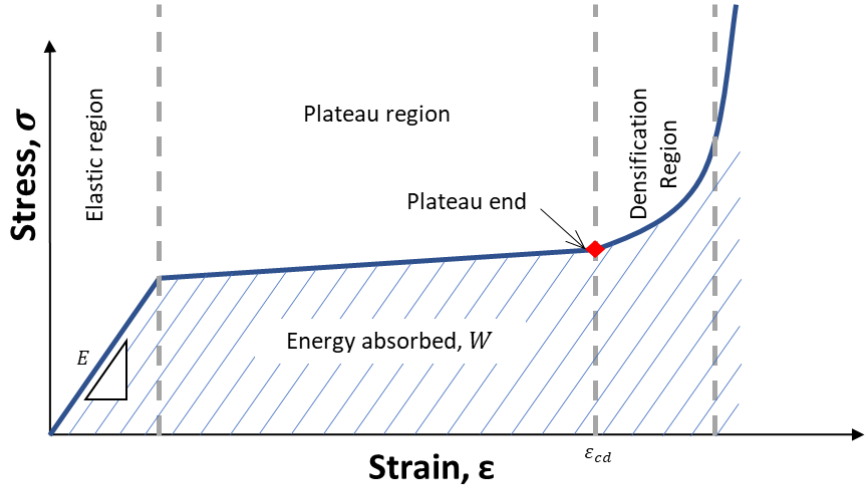

Figure 5: A schematic showing the typical stress-strain response for a lattice under compressive load, indicating the three main regions (reproduced from (Habib et al., 2018))

The energy absorbed per unit volume, $W$, by a lattice structure subjected to compressive loading up to a strain of $\varepsilon$ is determined by the area under the compressive stress-strain curve. According to ISO13314:2011, the energy absorption per unit volume is given by (Gibson, 2018; Habib et al., 2018):

$$W = \int_0^\varepsilon \sigma(\varepsilon) d\varepsilon \tag{7}$$

where $\sigma(\varepsilon)$ represents the stress the corresponding strain at $\varepsilon$. The total energy absorbed for each of the structures was calculated at the end of the plateau region. The total energy absorption was basically the total area under each curve for each strain intervals and was calculated using the trapezoidal rule as a numerical method.

The plateau region is the main energy absorbing part in the stress-strain curve of lattice structures. To identify the condition of the optimal energy absorption of foams, and cellular solids, the following energy absorption efficiency parameter, $\eta$, is introduced as (Q. M. Li et al., 2006; Yin et al., 2023):

$$\eta(\varepsilon) = \frac{\int_0^\varepsilon \sigma(\varepsilon) d\varepsilon}{\sigma(\varepsilon)} \tag{8}$$

As mentioned previously, energy absorption of lattice structures is most effective up to the end of the plateau region, or the onset of densification, after which the stress rises rapidly at the densification strain, with little to no improvement in energy absorption (Habib et al., 2018). To calculate the plateau end point or onset of densification strain, $\varepsilon_{cd}$, at which the energy absorption reaches its maximum, we can use (Q. M. Li et al., 2006):

$$\left.\frac{d\eta(\varepsilon)}{d\varepsilon}\right|_{\varepsilon=\varepsilon_{cd}} = 0 \qquad (9)$$

The above densification and energy absorption characteristics were computed for all lattices structures studied.

## 3 Results

### 3.1 Mesh sensitivity analysis

The simulation results for three node spacings for each of the gyroid and dual-lattice structures are presented in Figure 6. The mesh sensitivity analysis revealed that a node spacing of 13 μm was sufficiently small to guarantee converged results, and reasonable calculation time as indicated in Table 2.

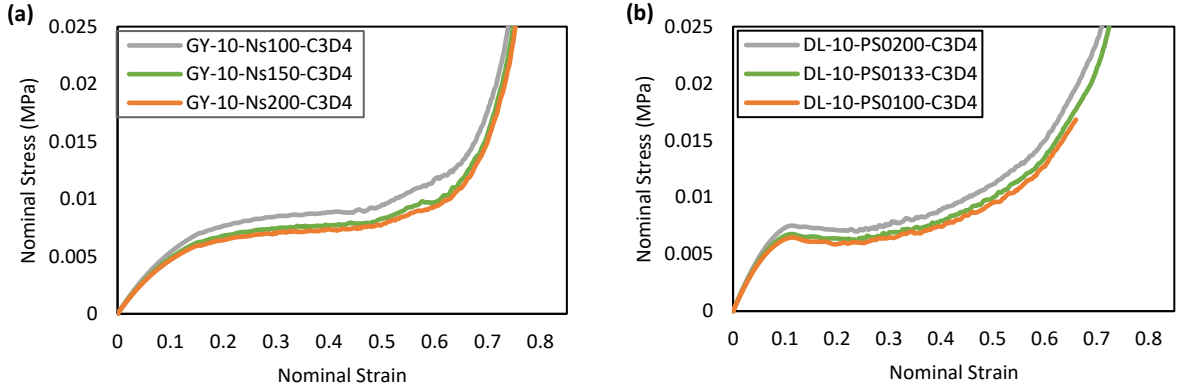

Figure 6: Convergence of stress-strain curve corresponding to three mesh seed sizes for 10% volume fraction models of (a) gyroid, and (b) dual-lattice structures. (Note that GY-10-Ns100 indicates a gyroid (GY), 10% volume fraction, with edge length that consisted of 100 nodes, whereas DL-10-PS0200 means a dual-lattice (DL) with 10% volume fraction and point spacing of 0.02 mm, that they result in the same mesh sizes).

### 3.2 Model performance

Figure 7 (a-b) shows the results from the uniaxial compression testing for both the gyroid and dual-lattice structures at 10% volume fraction from the experimental testing and computational simulation. In the gyroid structure, a uniform deformation was evident throughout the lattice under loading in Figure 7(a). At high levels of load, the central region of the lattice showed some outward convexity due to the buckling of the cells, giving the whole structure a barrel-like shape, clearly visible in Figure 7 (a). The layers collapsed due to the buckling of the connecting rods, and high stress bands in high deformation were observed in the computational models in these regions, in Figure 7 (a). In the dual-lattice structure, stress concentrations appeared in the buckling struts (edges) of the cells, as depicted in Figure 7 (b), and the deformation was not uniform, unlike the gyroid structure. The buckling of the long struts resulted in the entire structure bending in the middle, and asymmetric buckling deformation across the entire lattice, which was evident both the experimental and computational models.

Figure 7 (e-f) shows the stress-strain response of the mechanical compression tests on of the 3D printed gyroid and dual-lattice structures (n=3). According to the experimental data in both gyroid and dual-lattice cases, the initial bending of the struts resulted in linear elastic deformation to the structures. Due to the elastic buckling of the struts, a yield point occurred and there was an almost constant plateau stress with subsequent collapsing of the struts, causing increase in strain. When struts begin to come in contact with each other, there was a rapid increase in stress values, showing the densification region. As shown in Figure 7 (e-f), the stress-strain curve calculated from FEM based on the calibrated material showed good consistency with the mechanical test results. The FEM model captured both the deformation modes shown in Figure 7 (c-d) and the stress-strain curves in the FEM, that as shown in Figure 7 (e-f), are in good agreement with the experimental results.

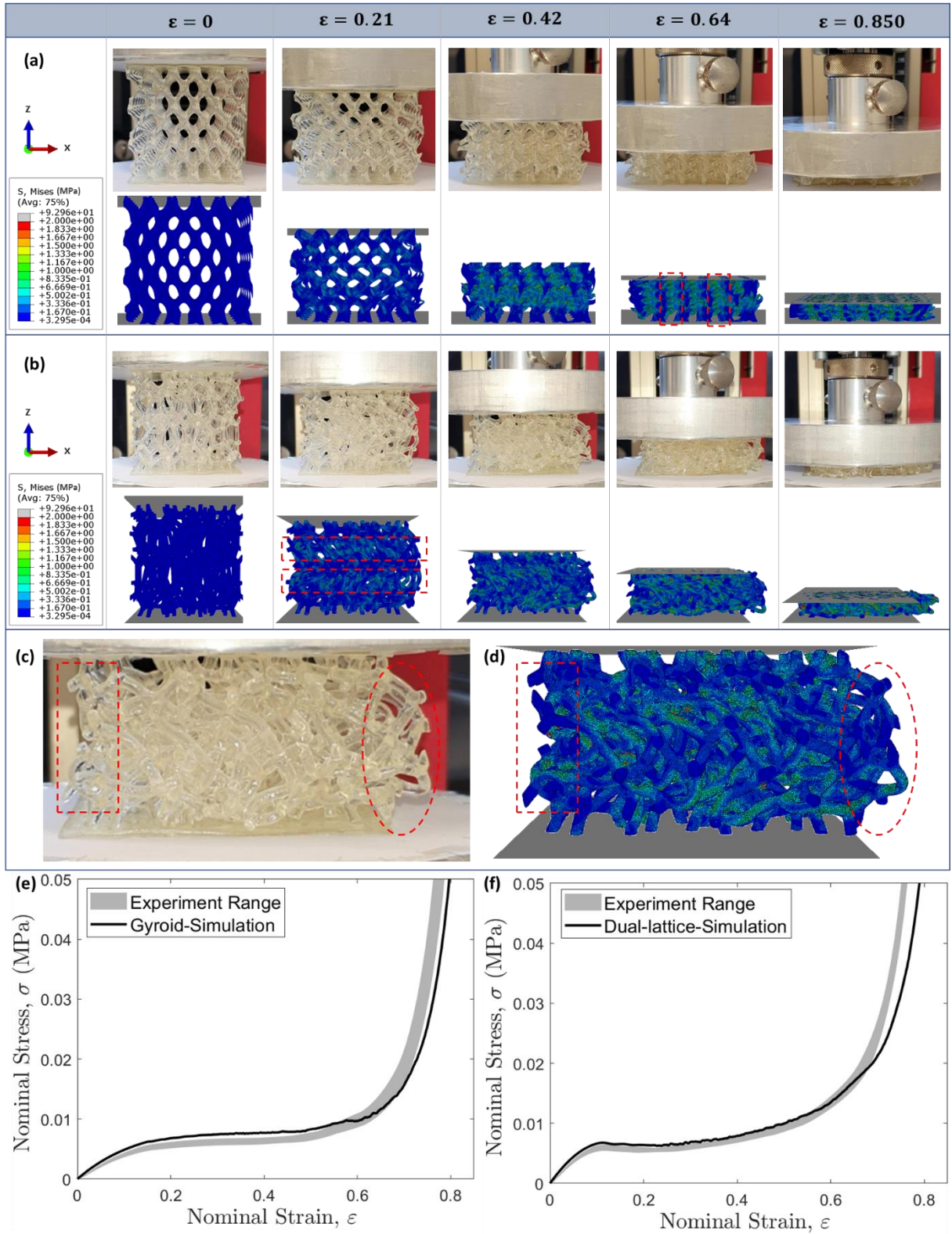

Figure 7: Compression response for the 10% volume fraction models of gyroid (a) and dual-lattice (b) structures: experimental and simulation results. Deformation mode captured by the FE analysis on 10% volume fraction of dual-lattice at strain $\varepsilon = 0.42$ (c, d). Validation of simulation and experiment results for the 10% volume fraction of the gyroid (e) and dual-lattice (f) structures.

### 3.3 Mechanical properties

#### 3.3.1 Stress-strain curves

Figure 8 shows the predicted stress-strain response for each of the lattice structures, for the parametric study on the volume fraction. Here, the mechanical response of the gyroid structures presents an initial linear region, followed by a flat plateau region, and an increasing stress in the densification region. For higher volume fractions, it was found that the length of the plateau region was reduced, and densification or stiffening occurred at lower strains. It is noticeable that the plateau region was barely evident in several of the structures at the highest volume fraction, with the structure proceeding almost directly to densification.

In Figure 8, the dual-lattice structure shows similar behaviour to the gyroid, but there is a drop in the stress before the structure proceeded to the plateau region. For higher volume fractions, the drop in stress before the plateau region was more severe. In general, during the linear region, and at high strains, the dual-lattice structure showed the highest stress values compared to the gyroid and spinodoid structures. The onset of densification occurred at smaller strains in the dual-lattice structure compared to the gyroid.

The spinodoid structure had the lowest mechanical response for all studied volume fractions, specifically at low volume fractions. Furthermore, the linear plateau and densification regions in the stress-strain behaviour of the spinodoid were not as distinct, or clearly observable, as for the two other structures. The transition between these three phases occurred more smoothly in the spinodoid structure. Unlike the repeating-unit lattices, in the spinodoid structure, collapsing of the struts was gradual, and did not occur at specific range of deformation. For higher volume fractions, the difference between the spinodoid and other two lattice in stress values at the same strain points became smaller.

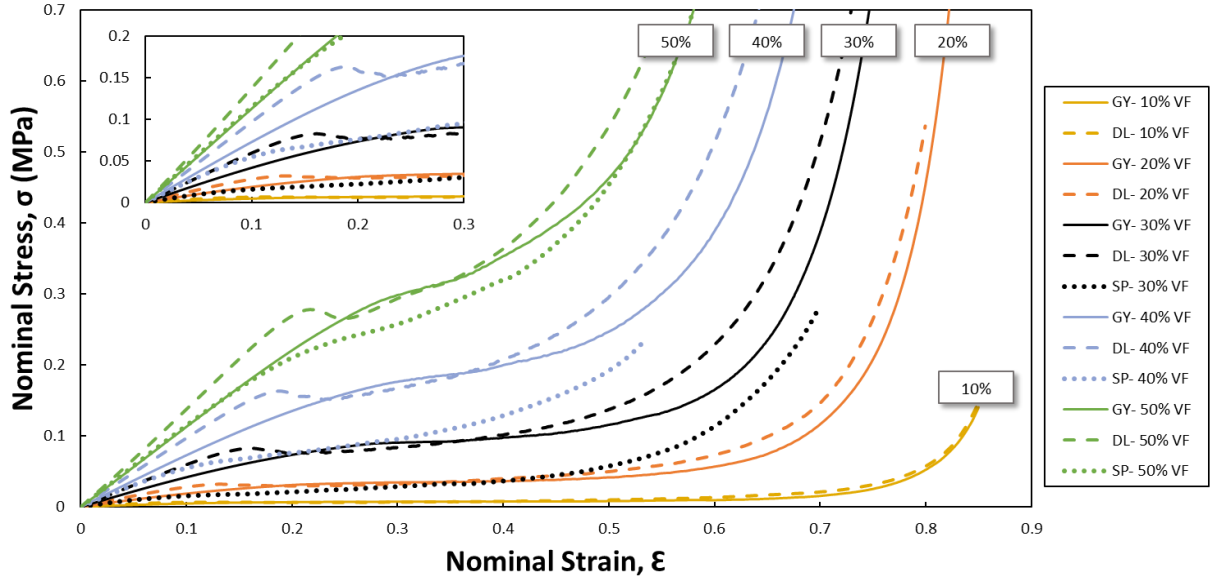

Figure 8: Stress-strain curves of three lattice types with variable volume fractions under compression loading from FE models.

Table 5 summarises the effective Young's moduli, $E$, yield stress, $\sigma_y$, plateau stress, $\sigma_{pl}$, onset strain of densification, $\varepsilon_{cd}$, as well as the energy absorption characteristics for each lattice and volume fraction. Consistent with our previous work (Vafaeefar et al., 2023), the dual lattice had the highest Young's modulus and yield stress, compared to the gyroid and spinodoid structures for the same volume fraction. The plateau stress ($\sigma_{pl}$) was similar for the gyroid and the dual-lattice structures, but the spinodoid structure showed lower plateau stresses. The strain at the onset of the densification ($\varepsilon_{cd}$) was lower than the theoretical densification strain ($\varepsilon_d = 1 - Vf$), as the cell struts jammed together at a lower strain in reality (Habib et al., 2018).

Table 5 Energy absorption and compressive properties of the three lattices derived from FEA stress-strain diagrams.

| Type | Design volume fraction, VF (%) | Effective Young's modulus, E (MPa) | Yield Stress, $\sigma_y$ (Mpa) | Plateau stress, $\sigma_{pl}$ (Mpa) | $\varepsilon_d$ | At the plateau end | | | |
|---|---|---|---|---|---|---|---|---|---|
| | | | | | | Strain, $\varepsilon_{cd}$ | Stress, $\sigma$ (MPa) | Energy absorbed, W (mJ/mm$^3$) | Efficiency, $\eta$ (%) |
| Gyroid- 10% VF | 10% | 0.064 | 0.002 | 0.007 | 0.89 | 0.609 | 0.010 | 0.004 | 42.3% |
| Gyroid- 20% VF | 20% | 0.221 | 0.009 | 0.030 | 0.80 | 0.573 | 0.051 | 0.018 | 35.0% |
| Gyroid- 30% VF | 30% | 0.453 | 0.026 | 0.091 | 0.70 | 0.548 | 0.131 | 0.042 | 32.2% |
| Gyroid- 40% VF | 40% | 0.749 | 0.065 | 0.162 | 0.60 | 0.521 | 0.266 | 0.077 | 28.8% |
| Gyroid- 50% VF | 50% | 1.118 | 0.215 | 0.330 | 0.50 | 0.514 | 0.487 | 0.129 | 26.5% |
| Spinodoid- 30% VF | 30% | 0.225 | 0.007 | 0.023 | 0.70 | 0.441 | 0.043 | 0.011 | 25.3% |
| Spinodoid- 40% VF | 40% | 0.645 | 0.029 | 0.089 | 0.60 | 0.471 | 0.182 | 0.046 | 25.2% |
| Spinodoid- 50% VF | 50% | 1.155 | 0.138 | 0.259 | 0.50 | 0.473 | 0.403 | 0.107 | 26.6% |
| Dual-lattice- 10% VF | 10% | 0.102 | 0.003 | 0.007 | 0.90 | 0.580 | 0.012 | 0.004 | 34.0% |
| Dual-lattice- 20% VF | 20% | 0.327 | 0.015 | 0.032 | 0.80 | 0.510 | 0.050 | 0.016 | 31.8% |
| Dual-lattice- 30% VF | 30% | 0.618 | 0.051 | 0.093 | 0.70 | 0.486 | 0.130 | 0.038 | 28.9% |
| Dual-lattice- 40% VF | 40% | 0.955 | 0.147 | 0.189 | 0.60 | 0.460 | 0.251 | 0.067 | 26.5% |
| Dual-lattice- 50% VF | 50% | 1.318 | 0.276 | 0.325 | 0.51 | 0.457 | 0.444 | 0.117 | 26.4% |

Figure 9 shows the stress distribution at 50% compressive strain at different volume fractions for the three types of lattices. Local stress distributions show higher stresses for higher volume fractions, which is a result of self-contact of the lattice components and contact stresses being induced in these high-volume fraction lattices. In Figure 9, comparing the same volume fraction of the gyroid and dual-lattice structures, the dual-lattice struts engaged in self-contact earlier than the gyroid structure, with pores closed at smaller strains. It is also notable from Figure 9 that the deformation and buckling modes of each lattice structure remained the same across all volume fractions. For example, the uniform deformation in the gyroid structure existed in all volume fractions, and in the dual-lattice structures, the convex bending on the $x$-axis occurred in all volume fractions. In the spinodoid structure, the highest stress concentrations took place on elements in contact, or, in thin connecting struts subjected to tension as the structure expanded laterally due to the external compressive load. On the boundaries, there were some "free struts", most evident for the spinodoid structure, which during high compression can become pushed out of the main structure, leaving them relatively unstressed.

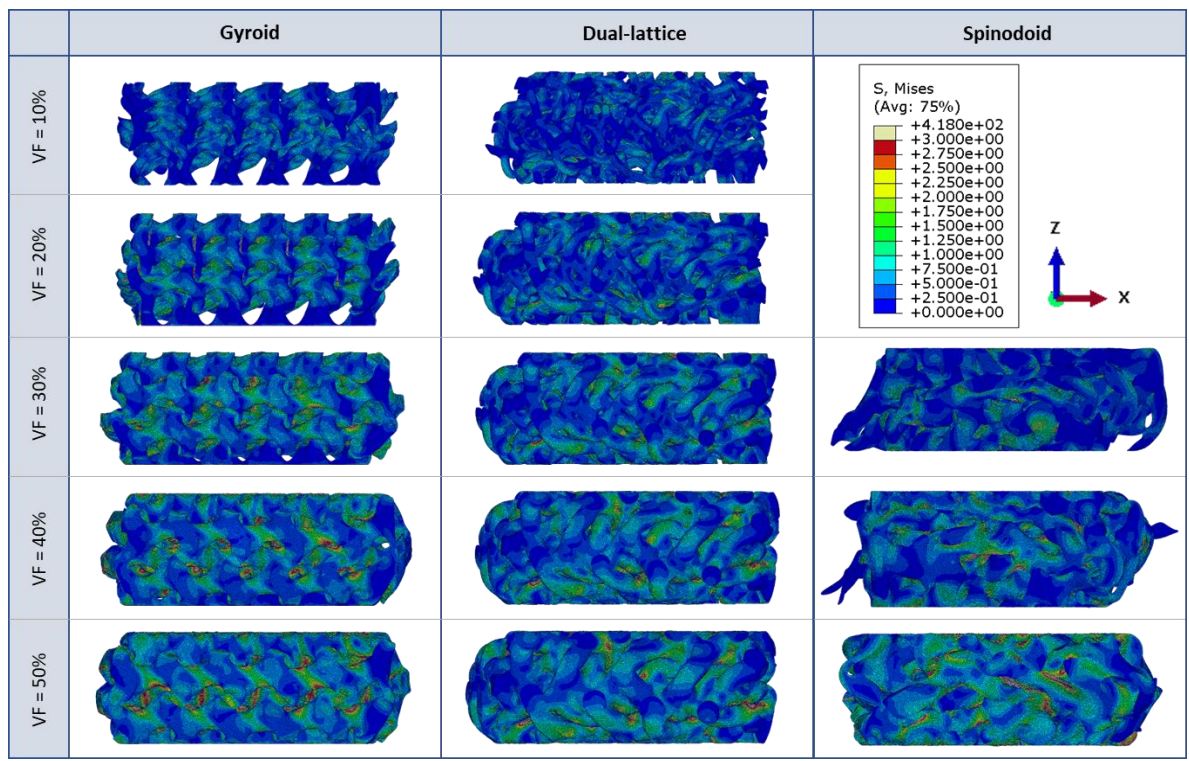

Figure 9: The deformed state and stress distributions for the gyroid, dual-lattice, and spinodoid structures at 50% compression for a range of volume fractions (VF).

*3.3.2 Energy absorption capability*

Figure 10 shows the cumulative energy absorption up to the start of densification regime calculated for each of the structures, as reported in Table 5, at each volume fraction. Here, the dual-lattice structure was the most energy absorbent compared to the other structures, for the same strain values, which is more evident for the higher volume fractions. The spinodoid structure was the least energy absorbent in for the same volume fraction, however the difference became smaller in VF = 50% volume fraction.

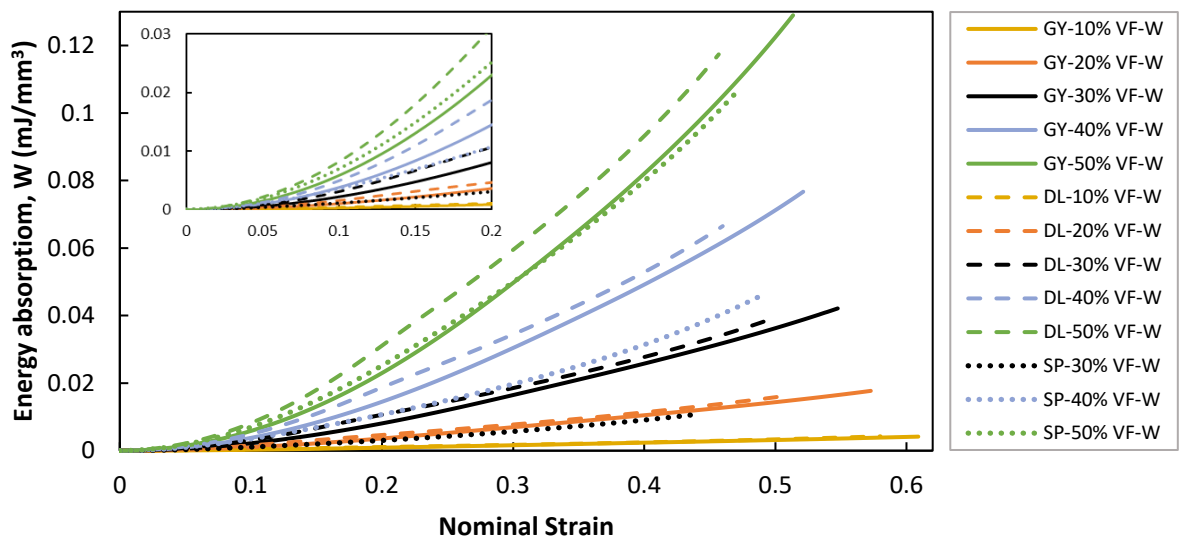

Figure 10: Energy absorption for three types of lattices, in different volume fractions up to their onset of densification strain.

Figure 11 shows the energy absorption efficiency for each lattice structure. The onset of densification strain was predicted by the efficiency curves in all structures. The energy absorption efficiency reduced with increased volume fraction, in the gyroid and dual-lattice structures. However, in the spinodoid structures, the onset of densification for each volume fraction was similar, and slightly increased with increasing volume fraction. As shown in Figure 11(a), the gyroid structure presented with higher ranges of efficiency for each volume fraction, compared with the two other structures. In Figure 10, although the absorbed energy was higher in dual-lattice structure at the same strains, the gyroid structures showed higher efficiencies, due to lower stress values, as presented in Figure 8.

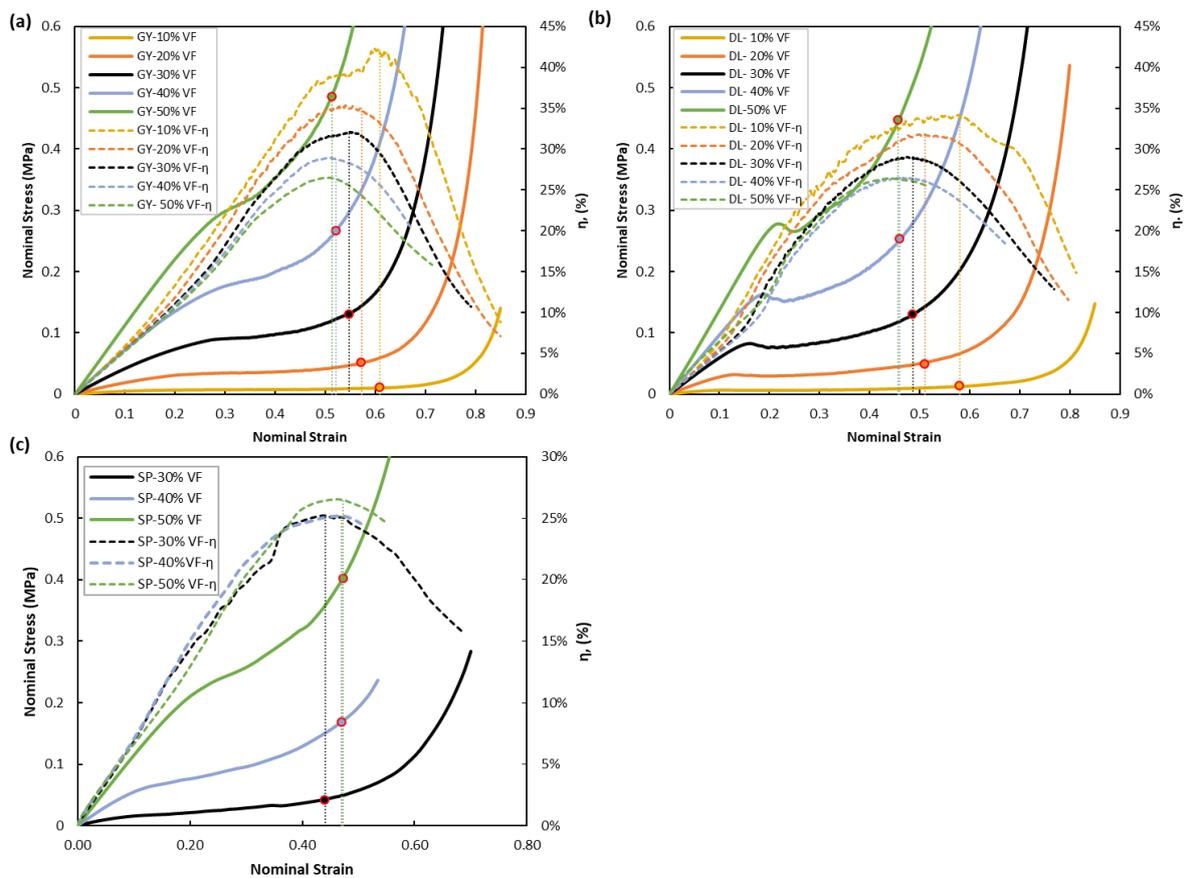

Figure 11: Energy absorption efficiency-strain curves of the gyroid (a), dual-lattice (b), and spinodoid (c) structures with different volume fractions, showing the onset of densification on stress-strain curves.

Table 5 summarises the mechanical and energy absorption characteristics for the three lattice structures and each volume fraction investigated. The results show that for each lattice structure, the energy absorption value at the onset of densification, and the plateau stress, increased as a function of the volume fraction. Table 5 also shows that the plateau stress was slightly higher for the same volume fraction in the dual-lattice structure and was the lowest in the spinodoid structure. Strain at the onset of densification, $\varepsilon_{cd}$, was the lowest for the dual-lattice structure at the same volume fractions. The total energy absorbed at the onset of densification was the highest for the gyroid structure since it densified at higher strain values for the same volume fraction.

*3.4 Microarchitectural study*

Table 6 reports the measured microarchitectural features of all the lattice structures at different volume fractions. Comparing the struts spacings showed that, at the same volume fraction, the spinodoid structure had the lowest mean spacing between the stuts, whereas the dual-lattice structure structs had the highest mean separation. The dual-lattice structure generally showed higher strut thickness compared to the gyroid and spinodoid structures, at the same volume fractions.

Table 6 Microstructural features measured for the three types of lattices in different volume fractions.

| Name | BV (mm$^3$) | TV (mm$^3$) | BV/TV | Tb.Th Mean (mm) | Tb.Sp Mean (mm) |
|---|---|---|---|---|---|
| Gyroid-10% VF | 0.853 | 8.000 | 0.107 | 0.082 | 0.307 |
| Gyroid-20% VF | 1.627 | 8.000 | 0.203 | 0.111 | 0.272 |
| Gyroid-30% VF | 2.415 | 8.000 | 0.302 | 0.136 | 0.242 |
| Gyroid-40% VF | 3.202 | 8.000 | 0.400 | 0.165 | 0.215 |
| Gyroid-50% VF | 4.023 | 8.000 | 0.503 | 0.188 | 0.187 |
| Spinodoid-30% VF | 2.410 | 8.000 | 0.301 | 0.145 | 0.226 |
| Spinodoid-40% VF | 3.221 | 8.000 | 0.403 | 0.162 | 0.202 |
| Spinodoid-50% VF | 4.011 | 8.000 | 0.501 | 0.182 | 0.183 |
| Dual-lattice-10% VF | 0.800 | 8.000 | 0.100 | 0.074 | 0.393 |
| Dual-lattice-20% VF | 1.622 | 8.000 | 0.203 | 0.112 | 0.349 |
| Dual-lattice-30% VF | 2.434 | 8.000 | 0.304 | 0.144 | 0.315 |
| Dual-lattice-40% VF | 3.216 | 8.000 | 0.402 | 0.173 | 0.288 |
| Dual-lattice-50% VF | 3.950 | 8.000 | 0.494 | 0.200 | 0.264 |

The relationship between energy absorption characteristics and measured microstructural features is presented in Figure 12. As evident in the stress-strain behaviour of the lattices in Figure 8, the length of the plateau region reduced with higher volume fractions, and occurred at higher stress values, and this is also reflected in Figure 12(a). In Figure 12(b), the onset of densification was reduced for the gyroid and dual-lattice structures with increase in volume fraction. However, the spinodoid structure showed a different trend, as it presented with an initial increase in the onset of densification strain as a function of volume fraction. In Figure 12(c), the energy absorption capacity increases with volume fraction for all the structures. Since the increasing volume fraction was applied by thickening the struts of the cellular solid, Figure 12(d) shows the same behaviour for the energy absorption and strut thickness (Tb.Th), for all the structures. Figure 12(e) shows that for the gyroid and dual-lattice structures, the energy absorption efficiency decreased with increased volume fraction, while energy absorption efficiencies in the spinodoid structure showed no noticeable differences.

In the gyroid and dual-lattice structures, as the volume fraction increases, the separation between the struts becomes smaller. As a result, the struts collapse and touch each other at a lower strain. This is reflected in Figure 12(f) showing the onset of densification strain vs. the strut's separation. This confirms that densification occurs at lower strains, as the porosity, and space between struts, is decreased. However, the spinodoid structure did not follow this behaviour, and shows no change or increase in onset of densification with reducing the mean strut separation.

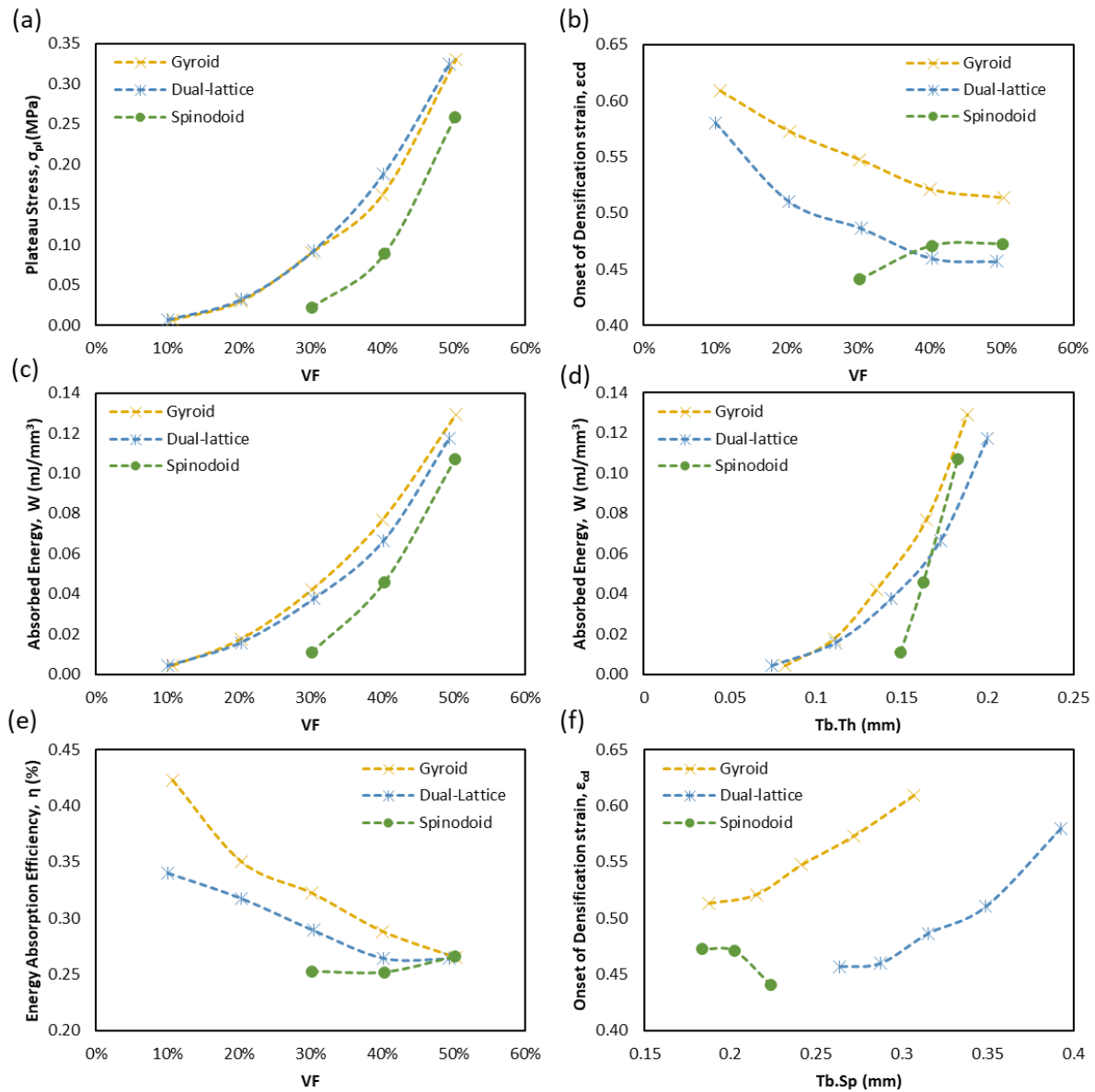

Figure 12 Energy absorption characteristics, W, η, $\sigma_{pl}$, and $\varepsilon_{cd}$ for each lattice structure, as functions of the microstructural parameters, VF, Tb.Th, and Tb.Sp.

### 3.5  *Dominant deformation and scaling law analysis*

Stronger nodal connectivity typically results in stretch-dominated behaviour (Kadkhodapour et al., 2014; Kaur et al., 2017). Nodal connectivity for each of the structures has been reported in our previous study (Vafaeefar et al., 2023). Unlike the gyroid unit cell with 3 edges connecting at each joint, the dual-lattice unit cell has a nodal connectivity of 4, demonstrating less bending dominant behaviour compared to the gyroid. The spinodoid structure has a mixed mode nodal connectivity from 3 to 5, with a majority of 3 nodal connectivity, that cannot be characterized based on its nodal connectivity.

Figure 13 shows the fitted power-law equations for the normalized Young's modulus vs the density ratio, for all the structure types. The gyroid and dual-lattice structure show good fit, and they follow the power-law equation. For the gyroid structure $n = 1.79$, and for the dual-lattice structure, $n = 1.53$, $c$ factor is close to 1 for the two lattices. Based on the power-law analysis (Ashby, 2006; Gibson & Ashby, 1997), the gyroid structure is more bending-dominated as the $n$ value is closer to 2 compared to the dual-lattice structure. However, the spinodoid structure does not show a good fit for the power-law equation, and there are large deviations in the FE points from the power-law fitted curve, indicating that spinodoid structure behaviour is not predictable with the power-fit law.

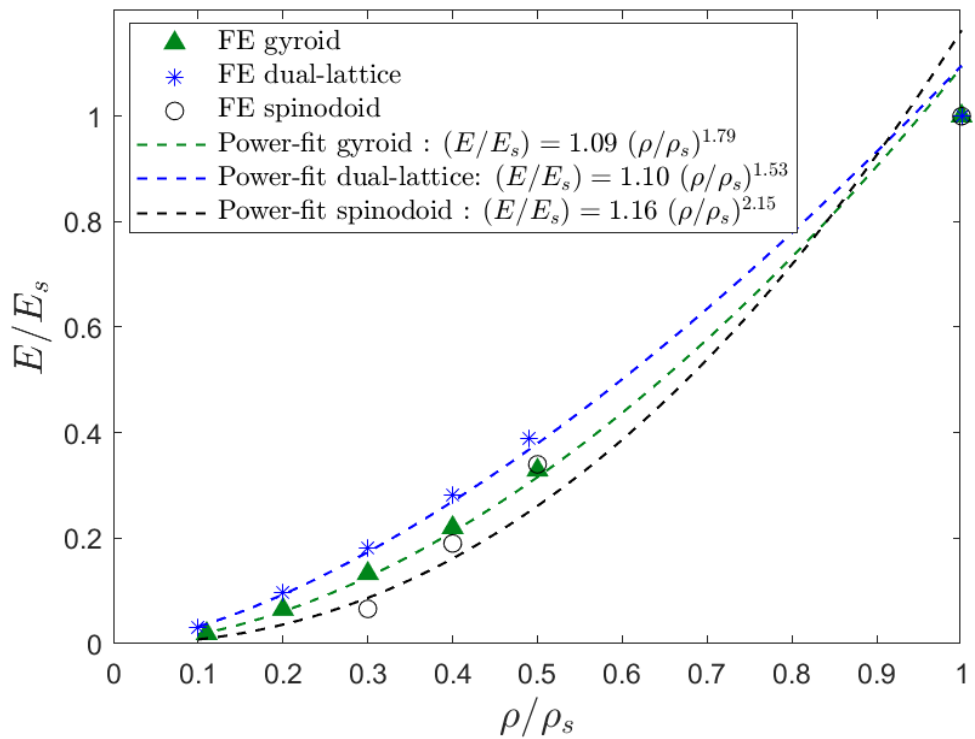

Figure 13 Normalized Young's modulus versus volume fraction and fitted power-law equations for the gyroid, dual-lattice, and spinodoid structures.

Based on the stress strain curve in Figure 8, in the gyroid structure the yield point follows a flat stress-strain response. Whereas the dual-lattice structure showed an early collapse, with the initial yield is followed by elastic buckling of the struts, which results in post-yield stress drop. The spinodoid structure deviates more from an ideal flat plateau region (Ashby, 2006; Deshpande et al., 2001).

## 4 Discussion

In this study, we considered three types of biomimetic lattice structures inspired by trabecular bone microarchitecture including the gyroid, dual-lattice and spinodoid structures (Vafaeefar et al., 2023).

Cubic samples for the gyroid and dual lattice were 3D printed, using a highly flexible elastic resin, and for a volume fraction of 10% volume fraction, and subjected to high strain compression testing (85%). Finite element (FE) models were created for all structures, and a range of volume fractions (10-50%), and were used to study the mechanical performance, energy absorption characteristics, and deformation modes of each structure type as a function of volume fraction. Both the experimental and simulated stress-strain curves clearly showed the typical initial linear region, the plateau region, and the densification regions characteristic of lattice structures. Based on the FE results, as the volume fraction increases, the energy absorption capacity of the lattices increased, and the plateau region shortened. The gyroid and dual-lattice structures showed a higher energy absorption capacity compared to the spinodoid structure, especially at lower volume fractions. The gyroid structure, presented with a higher energy absorption efficiency, and a higher onset of densification strain, compared to the two other structures at the same volume fraction. Furthermore, because of the bending-dominant deformation mechanism of the gyroid structure, as also indicated in other studies (Ataee et al., 2018; Lu et al., 2021; Wu et al., 2022), it is a suitable structure for energy absorption applications (X. Li et al., 2021).

The dual-lattice structure generally showed the highest capability to absorb strain energy, unlike the spinodoid structure which exhibited the lowest energy absorption. However, in terms of the energy absorption efficiency, the gyroid structure generally performed best, while the spinodoid structure presented with the lowest energy absorption efficiency. The energy absorption efficiency of the gyroid structure can perhaps be explained by its relatively low plateau stress combined with its relatively high onset of densification strain, $\varepsilon_{cd}$. In general, a structure that has a stress-strain curve with a long, flat plateau is desirable for energy absorption applications (Ashby, 2006; Deshpande et al., 2001). As a result, although the dual-lattice showed higher stiffness and strength, it is less suitable for energy absorption applications compared to the gyroid structure, at the same volume fraction. It also showed a post-yield stress drop, which happened due to the elastic buckling of the struts. In the spinodoid structure, transitions between the three regions, i.e., the initial elastic, the plateau, the densification regions, occurred more smoothly. In this structure, unlike the repeating- unit lattices, collapsing of the struts was gradual, and did not occur at a specific level of deformation. This behaviour could be due to the stochastic distribution of the struts and the spacing between them. Still, among the studied structures in this work, the gyroid structure is a better candidate for the energy absorption applications, due to its flat, long plateau region in the post-yield phase. Other studies have shown that the sheet-based (D. Li et al., 2019), and graded sheet-based (Zhou et al., 2020) gyroid performs even better than the strut-based models, that was considered in this study. However, here bone biomimetic structures were of interest, and hence these sheet-like variants were not included.

For all structures, microarchitectural features were assessed, and studied in relation to energy absorption characteristics. As shown in Figure 12 (a, c), for each structure, the plateau stress and energy absorption increased nonlinearly with volume fraction and, as expected, this effect is also seen as a function of strut thickness (see Figure 12 (d)). The energy absorption efficiency was observed to decrease with volume fraction in both the gyroid and dual-lattice structures, with a higher rate for the gyroid (Figure 12 (e)). However, increasing the volume fraction had a different effect on the spinodoid structure, as it resulted in a slight increase in onset of densification strain and energy absorption efficiency (Figure 12 (b, e)). The variations of these onset of densification are not strongly expressed, as the points are very close and there are some uncertainties regarding the exact hump of the energy efficiency curves. However, this different behaviour in the spinodoid structure might be explainable by poorly connected struts within the structure. These poorly connected struts close to the boundary in the spinodoid structure, moved out of the outlined boundaries during compression, as depicted in Figure 9 for 40% volume fraction. These protruding material parts were not engaged in the densification process. Therefore, this may explain why the densification occurred later in this structure (Figure 12 (b, f)). Comparing the deformation behaviour in the three types of lattices in Figure 9, the gyroid structure experiences the most uniform deformation, i.e., a spatially uniform type of elastic buckling in the thin struts connecting the layers of unit cells. In the dual-lattice structure, such buckling is not uniform, and regions of high stress were observed in the connecting joints (Figure 7). The spinodoid structure presented low load bearing capacity, that might be explained by the presence of the poorly connected and non-supported struts.

The gyroid is generally recognized as a bending dominant structure (Ataee et al., 2018; Lu et al., 2021; Wu et al., 2022). Considering the nodal connectivity of the skeletonised frame models of the structures (Vafaeefar et al., 2023), higher nodal connectivity in the dual-lattice, may explain its stretch-dominant behaviour with a post-yield softening in its stress-strain curve, and higher elastic modulus. The spinodoid structure has a non-uniform nodal connectivity; however, solely considering nodal connectivity, e.g., using the nodal connectivity criterion, is not a precise reference due to the lack of consideration of mass distribution, loading condition, strut thickness and material properties (Kadkhodapour et al., 2014; Keshavarzan et al., 2020). Hence, a power-law analysis was also considered for the structures. The scaling law analysis on the elastic modulus of all the structure types shown in Figure 13, confirmed that in the gyroid structure the $n = 1.79$ power is closer to 2, compared to the dual-lattice with $n = 1.53$, which is indicative of a more bending-dominant behaviour, and which makes it most desirable for energy absorption applications (Ashby, 2006). The spinodoid structure did not show a good fit based on the scaling law analysis, and its energy absorption behaviour is predictably not as good as the two other lattices.

This study directly compared the response of a computational model and experimental uniaxial compression test of the printed lattices at 10% volume fraction, which generally showed a good correlation based on the fitted parameters. However, it was found that the computational model response was sensitive to several aspects. Firstly, it was found that uniaxial compressive response of the samples was quite sensitive to the calibrated $D_1$ factor. This parameter is related to the compressibility of the material and, while a range of $D_1$ values were found to adequately capture the uniaxial tensile response (Bernini et al., 2023), it was determined that $D_1 = 0.25$ MPa$^{-1}$ (equivalent to a Poisson ratio of 0.43) provided the best fit for the uniaxial compression test data on the 3D printed structures. Furthermore, it was generally found that the FE results at high strain values slightly under-predicted the resulting stress values. This was mostly likely explained by the fact that in the FE model the buckled structure was not fully supported by the top and bottom rigid plates (Figure 9(b)). Therefore, part of the applied force, by the pushed-out elements, was not transferred to the plate. However, the support plates in the compression set-up are big enough to support the whole deformed cross section of the structure. As a result, the reaction force calculated in the FE model was lower than the experiment. In the gyroid structure, due to its more uniform collapse, which remained within its original outline, we didn't observe this difference in the FE simulation and experimental data. Furthermore, the FE response of the structures was a function of the mesh type and size. To overcome the mesh dependency, a detailed mesh study was performed, as shown in Figure 6. Due to our constraints in the computational cost, such as the maximum runtime with the cluster we used, 4-node tetrahedral elements (C3D4) were more practical to use in this study. Comparison of the FE results with higher order elements (C3D10M), with higher degrees of freedom, the computational cost increased noticeably, and these element types underwent excessive distortion and were not suitable for these large compression simulations. Finally, the experimental compressive response, for the 3D printed 10% volume fraction gyroid and dual-lattice structures (Figure 7), showed good ductility, and no sign of fracture up to 85% strain. However, there may have been some minor local damage-based processes taking place within the structure at these high strain levels, although did not manifest to any great extend in the overall uniaxial response.

This study is subject to several limitations, which are discussed here. Energy absorption applications of lattice structures, like personal protective devices, may experience dynamic impact loading in real life. However, in this study dynamic effects were not considered, and instead the focus was on evaluation of quasi-static behaviour. According to (Gibson & Ashby, 1997), applied strain rate in this study is considered as a "low" strain rate. Since the material used for 3D printing was an elastic resin, we reduced the strain rate to minimise any viscoelastic behaviour from the samples. Also, for elastomeric foams, the moduli are almost independent from the strain rate and is more dependent on the temperature (Gibson & Ashby, 1997). Although the structures' stress-strain response changes under impact loading, the overall comparison of the structures in terms of their energy absorption characteristics is still comparable. Also, this study uses additive manufacturing of an elastic resin using SLA printing. This technique, like any other additive manufacturing techniques, requires specific angles and orientations of overhangs. Otherwise, the use of internal supports would be inevitable. The structures and resin used in the study meant that internal supports were required to enable printing of these porous structures. The internal supports added to the main structure were thin and easily recognisable. However, removing these internal supports was the limiting factor for preparing 3D printed samples of all the three structures. These supports were difficult to access, and remove completely, while not damaging the structure of interest with the cutting tools. In the case of the spinodoid structure, the struts were poorly connected to be compatible with the 3D-printer constraints, and lots of internal supports were needed, that not all of them were accessible. In higher volume fractions of the structures, the access channels to these internal supports were narrower and not accessible. Therefore, we couldn't prepare 3D printed structures for high volume fractions and compare the experimental and simulation results individually for each. The 3D printed samples in this study were limited to the 10% volume fractions of the gyroid and dual-lattice structures. With that said, a more advanced additive manufacturing technique that allows printing such porous structures without any internal support is suggested. However, our developed FE model showed to be reliable with a good fit of experiment data.

## 5  Conclusion

In conclusion, three types of lattice structures were investigated in terms of their mechanical behaviour, energy absorption characteristics, and deformation modes, under high compression deformations. Samples of 10% volume fraction gyroid and dual-lattice structures were 3D printed and subjected to compression testing. Finite element models were able to closely capture the experimental response. A range of finite element models, for various volume fractions, enabled the analysis of the effect of volume fraction in these structures. For each structure, the energy absorption increased with volume fraction, and the plateau stress decreased in a non-linear fashion. Results also showed that the dual-lattice structure absorbed more energy, and the spinodoid was the least energy absorbent structure, at the same volume fraction. However, the onset of densification strain was highest for the gyroid structure, and the plateau region was also longer in the gyroid structure. The energy absorption efficiency of the gyroid structure was also greater. Furthermore, although the dual-lattice presented with the highest overall strength, the gyroid structure appeared the most bending-dominant structure, and presented with the highest energy absorption efficiency, and hence appeared the most desirable for energy absorption applications.

# 6 Acknowledgment

This project has received funding from the European Research Council (ERC) under the EU's Horizon 2020 research and innovation program (Grant agreement No. 804108).

# 7 References

*ABAQUS Analysis User's Manual* (6.14). (2014). Dassault Systèmes Simulia Corp.

*ABAQUS Finite Element Analysis Software*. (2019). Dassault Systèmes Simulia Corp.

Afshar, M., Pourkamali Anaraki, A., & Montazerian, H. (2018). Compressive characteristics of radially graded porosity scaffolds architectured with minimal surfaces. *Materials Science and Engineering: C*, *92*, 254–267. https://doi.org/10.1016/J.MSEC.2018.06.051

Andrew, J. J., Schneider, J., Ubaid, J., Velmurugan, R., Gupta, N. K., & Kumar, S. (2021). Energy absorption characteristics of additively manufactured plate-lattices under low- velocity impact loading. *International Journal of Impact Engineering*, *149*, 103768. https://doi.org/10.1016/J.IJIMPENG.2020.103768

Ashby, M. F. (2006). The properties of foams and lattices. *Philosophical Transactions of the Royal Society A: Mathematical, Physical and Engineering Sciences*, *364*(1838), 15–30. https://doi.org/10.1098/rsta.2005.1678

Ataee, A., Li, Y., Fraser, D., Song, G., & Wen, C. (2018). Anisotropic Ti-6Al-4V gyroid scaffolds manufactured by electron beam melting (EBM) for bone implant applications. *Materials & Design*, *137*, 345–354. https://doi.org/10.1016/J.MATDES.2017.10.040

Bernini, M., Hellmuth, R., Dunlop, C., Ronan, W., & Vaughan, J. (2023). Recommendations for finite element modelling of nickel-titanium stents — Verification and validation activities. *PLoS ONE*, *18*(8), 1–34. https://doi.org/10.1371/journal.pone.0283492

Bian, Y., Yang, F., Zhang, S., Chen, M., & Song, Y. (2021). Similarities of the Mechanical Responses of Body-Centered Cubic Lattice Structures with Different Constituent Materials Under Compression. *JOM*, *74*. https://doi.org/10.1007/s11837-021-04926-1

Cahn, J. W. (1965). Phase separation by spinodal decomposition in isotropic systems. *The Journal of Chemical Physics*, *42*(1), 93–99. https://doi.org/10.1063/1.1695731

Callens, S. J. P., Tourolle né Betts, D. C., Müller, R., & Zadpoor, A. A. (2021). The local and global geometry of trabecular bone. *Acta Biomaterialia*, *130*(xxxx), 343–361. https://doi.org/10.1016/j.actbio.2021.06.013

Colabella, L., Cisilino, A. P., Häiat, G., & Kowalczyk, P. (2017). Mimetization of the elastic properties of cancellous bone via a parameterized cellular material. *Biomechanics and Modeling in Mechanobiology*, *16*(5), 1485–1502. https://doi.org/10.1007/s10237-017-0901-y

Deshpande, V. S., Ashby, M. F., & Fleck, N. A. (2001). Foam topology: bending versus stretching dominated architectures. *Acta Materialia*, *49*(6), 1035–1040. https://doi.org/10.1016/S1359-6454(00)00379-7

Doube, M., Kłosowski, M. M., Arganda-Carreras, I., Cordelières, F. P., Dougherty, R. P., Jackson, J. S., Schmid, B., Hutchinson, J. R., & Shefelbine, S. J. (2010). BoneJ: Free and extensible bone image analysis in ImageJ. *Bone*, *47*(6), 1076–1079. https://doi.org/https://doi.org/10.1016/j.bone.2010.08.023

Echeta, I., Feng, X., Dutton, B., Leach, R., & Piano, S. (2020). Review of defects in lattice structures manufactured by powder bed fusion. *International Journal of Advanced Manufacturing Technology*, *106*(5–6), 2649–2668. https://doi.org/10.1007/s00170-019-04753-4

Fan, H. L., Meng, F. H., & Yang, W. (2006). Mechanical behaviors and bending effects of carbon fiber reinforced lattice materials. *Arch Appl Mech*, *75*, 635–647. https://doi.org/10.1007/s00419-006-0032-x

Fan, Z., Huang, G., Lu, Y., Chen, Y., Zeng, F., & Lin, J. (2022). Full compression response of FG-based scaffolds with varying porosity via an effective numerical scheme. *International Journal of Mechanical Sciences*, *223*, 107294. https://doi.org/10.1016/J.IJMECSCI.2022.107294

Fu, J., Liu, Q., Liufu, K., Deng, Y., Fang, J., & Li, Q. (2019). Design of bionic-bamboo thin-walled structures for energy absorption. *Thin-Walled Structures*, *135*, 400–413. https://doi.org/10.1016/J.TWS.2018.10.003

Gibson, L. J. (2005). Biomechanics of cellular solids. *Journal of Biomechanics*, *38*(3), 377–399. https://doi.org/10.1016/j.jbiomech.2004.09.027

Gibson, L. J. (2018). *Cellular Solids*. APRIL 2003, 270–274.

Gibson, L. J., & Ashby, M. F. (1997). *Cellular Solids: Structure and properties* (Second Edi). Cambridge University Press.

Ha, Lakes, R. S., & Plesha, M. E. (2019). Cubic negative stiffness lattice structure for energy absorption: Numerical and experimental studies. *International Journal of Solids and Structures*, *178–179*, 127–135. https://doi.org/10.1016/J.IJSOLSTR.2019.06.024

Ha, N. S., & Lu, G. (2020). A review of recent research on bio-inspired structures and materials for energy absorption applications. *Composites Part B: Engineering*, *181*, 107496. https://doi.org/10.1016/J.COMPOSITESB.2019.107496

Habib, F. N., Iovenitti, P., Masood, S. H., & Nikzad, M. (2017). In-plane energy absorption evaluation of 3D printed polymeric honeycombs. *Virtual and Physical Prototyping*, *12*(2), 117–131. https://doi.org/10.1080/17452759.2017.1291354

Habib, F. N., Iovenitti, P., Masood, S. H., & Nikzad, M. (2018). Fabrication of polymeric lattice structures for optimum energy absorption using Multi Jet Fusion technology. *Materials & Design*, *155*, 86–98. https://doi.org/10.1016/J.MATDES.2018.05.059

Kadkhodapour, J., Montazerian, H., & Raeisi, S. (2014). Investigating internal architecture effect in plastic deformation and failure for TPMS-based scaffolds using simulation methods and experimental procedure. *Materials Science and Engineering: C*, *43*, 587–597. https://doi.org/10.1016/J.MSEC.2014.07.047

Kansara, H., Koh, G., Varghese, M., Luk, J. Z. X., Gomez, E. F., Kumar, S., Zhang, H., Martínez-Pañeda, E., & Tan, W. (2021). Data-driven modelling of scalable spinodoid structures for energy absorption. *ArXiv*.

Kaur, M., Min Han, S., & Soo Kim, W. (2017). Three-dimensionally printed cellular architecture materials: perspectives on fabrication, material advances, and applications. *MRS Communications*, *7*(1). https://doi.org/10.1557/mrc.2016.62

Keshavarzan, M., Kadkhodaei, M., Badrossamay, M., & Karamooz Ravari, M. R. (2020). Investigation on the failure mechanism of triply periodic minimal surface cellular structures fabricated by Vat photopolymerization additive manufacturing under compressive loadings. *Mechanics of Materials*, *140*, 103150. https://doi.org/10.1016/J.MECHMAT.2019.103150

Kirby, M., Morshed, A. H., Gomez, J., Xiao, P., Hu, Y., Guo, X. E., & Wang, X. (2020). Three-dimensional rendering of trabecular bone microarchitecture using a probabilistic approach. *Biomechanics and Modeling in Mechanobiology*, *19*(4), 1263–1281. https://doi.org/10.1007/s10237-020-01286-8

Kumar, S., Tan, S., Zheng, L., & Kochmann, D. M. (2020). Inverse-designed spinodoid metamaterials. *Npj Computational Materials*, *6*(1), 1–10. https://doi.org/10.1038/s41524-020-0341-6

Kwon, Y., Thornton, K., & Voorhees, P. W. (2009). The topology and morphology of bicontinuous interfaces during coarsening. *Epl*, *86*(4). https://doi.org/10.1209/0295-5075/86/46005

Lévy, B., & Bonneel, N. (2013). Variational Anisotropic Surface Meshing with Voronoi Parallel Linear Enumeration. In X. Jiao & J.-C. Weill (Eds.), *Proceedings of the 21st International Meshing Roundtable* (pp. 349–366). Springer Berlin Heidelberg.

Li, D., Liao, W., Dai, N., & Xie, Y. M. (2019). materials Comparison of Mechanical Properties and Energy Absorption of Sheet-Based and Strut-Based Gyroid Cellular Structures with Graded Densities. *Materials*, *12*(13), 2183. https://doi.org/10.3390/ma12132183

Li, Q. M., Magkiriadis, I., & Harrigan, J. J. (2006). Compressive Strain at the Onset of Densification of Cellular Solids. *Journal of Cellular Plastics*, *42*, 371–392. https://doi.org/10.1177/0021955X06063519

Li, X., Xiao, L., & Song, W. (2021). Compressive behavior of selective laser melting printed Gyroid structures under dynamic loading. *Additive Manufacturing*, *46*, 102054. https://doi.org/10.1016/J.ADDMA.2021.102054

Liu, X. S., Sajda, P., Saha, P. K., Wehrli, F. W., & Guo, X. E. (2006). Quantification of the roles of trabecular microarchitecture and trabecular type in determining the elastic modulus of human trabecular bone. *Journal of Bone and Mineral Research*, *21*(10), 1608–1617. https://doi.org/10.1359/jbmr.060716

Lu, C., Zhang, C., Wen, P., & Chen, F. (2021). Mechanical behavior of Al–Si10–Mg gyroid surface with variable topological parameters fabricated via laser powder bed fusion. *Journal of Materials Research and Technology*, *15*, 5650–5661. https://doi.org/10.1016/J.JMRT.2021.11.008

Ma, Z., Zhang, D. Z., Liu, F., Jiang, J., Zhao, M., & Zhang, T. (2020). Lattice structures of Cu-Cr-Zr copper alloy by selective laser melting: Microstructures, mechanical properties and energy absorption. *Materials & Design*, *187*, 108406. https://doi.org/10.1016/J.MATDES.2019.108406

Mangipudi, K. R., Epler, E., & Volkert, C. A. (2016). Topology-dependent scaling laws for the stiffness and strength of nanoporous gold. *Acta Materialia*, *119*, 115–122. https://doi.org/10.1016/j.actamat.2016.08.012

Maskery, I., Aboulkhair, N. T., Aremu, A. O., Tuck, C. J., & Ashcroft, I. A. (2017). Compressive failure modes and energy absorption in additively manufactured double gyroid lattices. *Additive Manufacturing*, *16*, 24–29. https://doi.org/10.1016/j.addma.2017.04.003

Moerman, M. K. (2018). GIBBON: The Geometry and Image-Based Bioengineering add-On. *Journal of Open Source Software*, *3*(22), 506. https://doi.org/10.21105/joss.00506

Mueller, J., Matlack, K. H., Shea, K., & Daraio, C. (2019). Energy Absorption Properties of Periodic and Stochastic 3D Lattice Materials. *Advanced Theory and Simulations*, *2*(10). https://doi.org/10.1002/ADTS.201900081

Padilla, F., GRIMAL, Q., & LAUGIER, P. (2008). Ultrasonic Propagation Through Trabecular Bone Modeled as a Random Medium. *Japanese Journal of Applied Physics*, *47*(5), 4220–4222 #2008. https://doi.org/10.1143/JJAP.47.4220

Peng, C., & Tran, P. (2020). Bioinspired functionally graded gyroid sandwich panel subjected to impulsive loadings. *Composites Part B: Engineering*, *188*, 107773. https://doi.org/10.1016/J.COMPOSITESB.2020.107773

Peng, C., Tran, P., Nguyen-Xuan, H., & Ferreira, A. J. M. (2020). Mechanical performance and fatigue life prediction of lattice structures: Parametric computational approach. *Composite Structures*, *235*, 111821. https://doi.org/10.1016/J.COMPSTRUCT.2019.111821

Rahman, H., Yarali, E., Zolfagharian, A., Serjouei, A., & Bodaghi, M. (2021). Energy absorption and mechanical performance of functionally graded soft–hard lattice structures. *Materials*, *14*(6). https://doi.org/10.3390/ma14061366

Rammohan, A. V., Lee, T., & Tan, V. B. C. (2015). A Novel Morphological Model of Trabecular Bone Based on the Gyroid. *International Journal of Applied Mechanics*, *07*(03), 1550048. https://doi.org/10.1142/S1758825115500489

Ruiz, O., Schouwenaars, R., Ramírez, E. I., Jacobo, V. H., & Ortiz, A. (2010). Analysis of the architecture and mechanical properties of cancellous bone using 2D voronoi cell based models. *WCE 2010 - World Congress on Engineering 2010*, *1*(June), 609–614.

Ruiz, O., Schouwenaars, R., Ramírez, E. I., Jacobo, V. H., & Ortiz, A. (2011). Effects of architecture, density and connectivity on the properties of trabecular bone: A two-dimensional, Voronoi cell based model study. *AIP Conference Proceedings*, *1394*(October), 77–89. https://doi.org/10.1063/1.3649938

Si, H. (2015). TetGen, a Delaunay-Based Quality Tetrahedral Mesh Generator. *ACM Trans. Math. Softw.*, *41*(2). https://doi.org/10.1145/2629697

Silva, M. J., & Gibson, L. J. (1997). Modeling the mechanical behavior of vertebral trabecular bone: Effects of age-related changes in microstructure. *Bone*, *21*(2), 191–199. https://doi.org/10.1016/S8756-3282(97)00100-2

Soyarslan, C., Bargmann, S., Pradas, M., & Weissmüller, J. (2018). 3D stochastic bicontinuous microstructures: Generation, topology and elasticity. *Acta Materialia*, *149*, 326–340. https://doi.org/10.1016/j.actamat.2018.01.005

Tarafdar, A., Liaghat, G., Ahmadi, H., Razmkhah, O., Chitsaz Charandabi, S., Rezaei Faraz, M., & Pedram, E. (2021). Quasi-static and low-velocity impact behavior of the bio-inspired hybrid Al/GFRP sandwich tube with hierarchical core: Experimental and numerical investigation. *Composite Structures*, *276*, 114567. https://doi.org/10.1016/J.COMPSTRUCT.2021.114567

Tsang, H. H., & Raza, S. (2018). Impact energy absorption of bio-inspired tubular sections with structural hierarchy. *Composite Structures*, *195*, 199–210. https://doi.org/10.1016/J.COMPSTRUCT.2018.04.057

Ufodike, C. O., Wang, H., Ahmed, M. F., Dolzyk, G., & Jung, S. (2021). Design and modeling of bamboo biomorphic structure for in-plane energy absorption improvement. *Materials & Design*, *205*, 109736. https://doi.org/10.1016/J.MATDES.2021.109736

Vafaeefar, M., Moerman, K. M., Kavousi, M., & Vaughan, T. J. (2023). A morphological, topological and mechanical investigation of gyroid, spinodoid and dual-lattice algorithms as structural models of trabecular bone. *Journal of the Mechanical Behavior of Biomedical Materials*, *138*(September 2022), 105584. https://doi.org/10.1016/j.jmbbm.2022.105584

Wang, Y., Ren, X., Chen, Z., Jiang, Y., Cao, X., Fang, S., Zhao, T., Li, Y., & Fang, D. (2020). Numerical and experimental studies on compressive behavior of Gyroid lattice cylindrical shells. *Materials & Design*, *186*, 108340. https://doi.org/10.1016/J.MATDES.2019.108340

Wu, S., Yang, L., Yang, X., Chen, P., Su, J., Wu, H., Liu, Z., Wang, H., Wang, C., Yan, C., & Shi, Y. (2022). Mechanical properties and energy absorption of AlSi10Mg Gyroid lattice structures fabricated by selective laser melting. *Smart Manufacturing*, *01*(01). https://doi.org/10.1142/S2737549821500010

Xiang, X., Zou, S., Ha, N. S., Lu, G., & Kong, I. (2020). Energy absorption of bio-inspired multi-layered graded foam-filled structures under axial crushing. *Composites Part B: Engineering*, *198*, 108216. https://doi.org/10.1016/J.COMPOSITESB.2020.108216

Yang, L., Mertens, R., Ferrucci, M., Yan, C., Shi, Y., & Yang, S. (2019). Continuous graded Gyroid cellular structures fabricated by selective laser melting: Design, manufacturing and mechanical properties. *Materials & Design*, *162*, 394–404. https://doi.org/10.1016/J.MATDES.2018.12.007

Yin, H., Zhang, W., Zhu, L., Meng, F., Liu, J., & Wen, G. (2023). Review on lattice structures for energy absorption properties. *Composite Structures*, *304*, 116397. https://doi.org/10.1016/J.COMPSTRUCT.2022.116397

Yu, S., Sun, J., & Bai, J. (2019). Investigation of functionally graded TPMS structures fabricated by additive manufacturing. *Materials & Design*, *182*, 108021. https://doi.org/10.1016/J.MATDES.2019.108021

Zhang, C., Jin, J., He, M., & Yang, L. (2022). Compressive Mechanics and Hyperelasticity of Ni-Ti Lattice Structures Fabricated by Selective Laser Melting. *Crystals*, *12*(3), 408. https://doi.org/10.3390/cryst12030408

Zhang, Chen, X., Sun, Y., Yang, J., Chen, R., Xiong, Y., Hou, W., & Bai, L. (2022). Design of a biomimetic graded TPMS scaffold with quantitatively adjustable pore size. *Materials & Design*, *218*, 110665. https://doi.org/10.1016/J.MATDES.2022.110665

Zhang, X. Y., Yan, X. C., Fang, G., & Liu, M. (2020). Biomechanical influence of structural variation strategies on functionally graded scaffolds constructed with triply periodic minimal surface. *Additive Manufacturing*, *32*, 101015. https://doi.org/10.1016/J.ADDMA.2019.101015

Zhou, H., Zhao, M., Ma, Z., Zhang, D. Z., & Fu, G. (2020). Sheet and network based functionally graded lattice structures manufactured by selective laser melting: Design, mechanical properties, and simulation. *International Journal of Mechanical Sciences*, *175*. https://doi.org/10.1016/j.ijmecsci.2020.105480